\documentclass[review,number,sort&compress,3p,times]{elsarticle}	
 \usepackage{lineno}
 \usepackage{color}

\usepackage{amssymb}
\usepackage{subfigure}
\begin{document}
\begin{frontmatter}
\title{Study of nanodiamond photocathodes for MPGD-based detectors of single photons }

\author[f]{F.~M.~Brunbauer}
\author[a]{C.~Chatterjee}
\author[d]{G.~Cicala}
\author[e]{A.~Cicuttin}
\author[e]{M.~L.~Crespo}
\author[b]{D.~D`Ago}
\author[a]{S.~Dalla~Torre}
\author[a]{S.~Dasgupta}
\author[a]{M.~Gregori}
\author[a]{S.~Levorato}
\author[c]{T.~Ligonzo}
\author[f,g]{M.~Lisowska}
\author[d]{M.~S.~Leone}
\author[e]{R.~Rai}
\author[f]{L.~Ropelewski}
\author[a]{F.~Tessarotto}
\author[a]{Triloki\corref{cor}}
\ead{ttriloki@cern.ch}
\author[c]{A.~Valentini}
\author[d]{L.~Velardi}

\cortext[cor]{Corresponding author}

\affiliation[a]{INFN Trieste  Trieste Italy}
\affiliation[b]{University of Trieste and INFN Trieste, Trieste, Italy}
\affiliation[c]{University Aldo Moro of Bari and INFN Bari, Bari, Italy}
\affiliation[d]{CNR-ISTP and INFN  Bari, Bari, Italy}
\affiliation[e]{Abdus Salam ICTP, Trieste, Italy and INFN Trieste, Trieste, Italy}
\affiliation[f]{European Organization for Nuclear Research (CERN), CH-1211 Geneve 23, Switzerland}
\affiliation[g]{Wrocław University of Science and Technology, Wybrzeże Wyspiańskiego 27, 50-370 Wrocław, Poland}


\begin{abstract}
 The proposed new Electron-Ion Collider poses a technical and intellectual challenge for the detector design to accommodate the long-term diverse physics goals envisaged by the program. This requires a 4$\pi$ detector system capable of reconstructing the energy and momentum of final state particles with high precision. The Electron-Ion Collider also requires identification of particles of different masses over a wide momentum range.

\par
A diverse spectrum of Particle IDentification detectors has been proposed. Of the four types of detectors for hadron identification, three are based on Ring Imaging Cherenkov Counter technologies, and one is realized by the Time of Flight method. The quest for a novel photocathode, sensitive in the far vacuum ultra violet wavelength range and more robust than cesium iodide, motivated an R\&D programme to explore nano-diamond (ND) based photocathodes, started by a collaboration between INFN and CNR Bari and INFN Trieste. Systematic measurements of the photo emission in different Ar-$CH_{4}$ and  Ar-$CO_{2}$ gas mixtures with various types of ND powders and Hydrogenated ND (H-ND) powders are reported. A first study of the response of THGEMs coated with different photocathode materials is presented. 
\par
The progress of this R\&D programme and the results obtained so far by these exploratory studies are described.
  
\end{abstract}

\end{frontmatter}

\section{Introduction}
\label{sec:intro}
The EPIC collaboration ~\cite{EPIC} which is the result of merging two proto-collaborations, Electron Ion Collider (EIC) Comprehensive Chromodynamics Experiment (ECCE) and A Totally Hermetic Electron-Nucleus Apparatus (ATHENA) existed from the start of the EIC. The future EIC~\cite{EIC} is the facility dedicated to understanding Quantum ChromoDynamics (QCD), including the elusive non-perturbative effects and the answer to key questions, pending since long. Among them: the origin of nucleon mass and spin  and the properties of dense gluon systems. The experimental activity at EIC requires  efficient hadron Particle IDentification (PID) in a wide momentum range, including the challenging scope of hadron PID at high momenta, namely larger than $6$-$8~GeV/c$. A gaseous Ring Imaging CHerenkov (RICH) is the only possible choice for this specific task. The number of Cherenkov photons generated in a light radiator is limited. In spectrometer setups,  these number of photons is recovered by using long radiators. The compact design of the experimental setup at the EIC collider imposes limitations on the radiator length, requiring a dedicated strategy. In the far ultraviolet (UV) spectral region ($\sim120$~nm), the number of generated Cherenkov photons is larger, according to the Frank-Tamm distribution~\cite{Frank:1937fk}. This suggests the detection of photons in the very far UV range. The standard fused-silica windows are opaque for wavelengths below 165~nm. Therefore, a windowless RICH is a potential option, implying the use of gaseous photon detectors operated with the radiator gas itself~\cite{windowless-RICH}.
\par 
The MicroPattern Gaseous Detector (MPGD)-based Photon Detectors (PD) \cite{MPGD} have recently been demonstrated as effective devices~\cite{PM18} for the detection of single photon in  Cherenkov imaging counters. These PDs are composed of a hybrid structure, where two layers of THick GEM (THGEM) multipliers~\cite{thgem} are followed by a MICRO-MEsh GAseous Structure (MICROMEGAS)~\cite{mm} stage; the top layer of the first THGEM is coated with a reflective CsI PhotoCathode (PC).
\par 
CsI PCs are, so far, the only available option for gaseous detectors of single photons, thanks to the relatively high work function that makes CsI more robust than other PC materials commonly
used in vacuum-based detectors. CsI has high Quantum Efficiency (QE) in the far UV spectral region. In spite of its widespread use and successful applications~\cite{RD26},  it presents  problematic aspects.  It is hygroscopic: the absorbed water vapour splits the CsI molecule, causing a degradation in QE~\cite{NIMA_695_2012_279}. Therefore,  the handling of CsI PC is a very delicate operation. QE degradation also appears after intense ion  bombardment, when the integrated charge is $\sim 1~mC/cm^{2}$~\cite{NIMA_574_2007_28} or larger. In gaseous detectors, ion bombardment of the cathode is caused by the ion avalanche produced in the multiplication process. The fraction of ions reaching the cathode depends on the detector architecture. In recent years, MPGD schemes with enhanced ion blocking capability have been developed~\cite{PM18, ibf-blocking}.
\par
The quest for an alternative UV sensitive photocathode overcoming these limitations is an important motivation for the R\&D programme for future gaseous PDs. In this article, we  present the preliminary results on the photon extraction efficiency in different $Ar/CH_{4}$ gas mixture and a comparison between H-ND coated THGEM detectors with a standard non coated THGEM.


\section{Nanodiamond based PC as an alternative of CsI PC}
The high QE value of CsI photocathodes is related to its low electron affinity ($0.1~eV$) and wide band gap ($6.2~eV$) ~\cite{JAP_77_1995_2138}. The ND particles have a comparable band gap of $5.5~eV$ and low electron affinity of $0.35$-$0.50~eV$, and they exhibit chemical inertness and radiation hardness. ND hydrogenation lowers the electron affinity to -$1.27~eV$. The Negative Electron Affinity (NEA) allows an efficient escape into vacuum of the generated photo electrons without an energy barrier at the surface~\cite{NDRep-1}. A novel ND hydrogenation and coating procedure, developed in Bari~\cite{NDRep-1, NDreport}, provides high and stable QE. 
In literature H-ND is reported by Velardi et al., to have 22\% QE at 146 nm , while QE of CsI at 146 nm is reported by Rabus at al., to be ~$\sim {45}\%$ at ${60}^{0}C$  and~$\sim {40}\%$ at ${25}^{0}C$ ~\cite{NDRep-1, NIMA_438_1999_94}.


\section{The R\&D activity}

\subsection{ND hydrogenation and coating procedure}
\par
Three types of ND powders have been used in 2021 for the set of measurements presented in this article. Two types have an average grain size of 250 nm and were produced by Diamonds \& Tools (D\&T) srl and Element 6 (E6); the third type, boron doped (BDD) produced by SOMEBETTER : ChangSha 3 Better Ultra-hard     materials Co.,LTD
 has an average grain size of 500 nm. 
\par
The standard procedure of hydrogenation of ND powder photocathodes is performed by using the MicroWave Plasma Enhanced Chemical Vapor Deposition (MWPECVD) technique. For the treatment in microwave (mw) $H_{2}$ plasma, 30 mg of ND powder was placed in a tungsten boat (overall length 32 mm, through 12 mm long × 5 mm wide × 1 mm deep, from Agar Scientific Ltd) positioned on a heatable substrate holder of an ASTeX-type reactor~\cite{JPCM_21_2009_16, Pat_Spec_5_501_740} evacuated to a base pressure better than $7\times{10}^{-7}$ mbar. The powder was heated to $650~^{0}$C using an external radiative heater (via a Proportional-Integral-Derivative feedback control system), then $H_{2}$ gas was flowed in the chamber at 200 sccm, the pressure and the mw power were maintained at 50 mbar and 1250 W, respectively. The heating due to the mw power increases further the temperature of the powders up to $1138~^{0}$C as determined by a dual wavelength ($\lambda _{1}$ = 2.1 $\mu m$ and $\lambda _{2}$ =2.4 $\mu m$) infrared pyrometer (Williamson Pro 9240). After 1 h of $H_{2}$ plasma exposure, the hydrogenated powder was cooled down to room temperature under high vacuum.
\par
This procedure cannot be used for THGEMs which are made of fiberglass, which does not tolerate temperatures above $180~^{0}$C.
This limitation is overcome by the novel and low-cost technique developed at INFN Bari  ~\cite{coating-I, coating-II}. 
\par
The three H-ND and the non hydrogenated ND powders were separately mixed with deionized water in one to one proportion and sonicated for 60 minutes. The solutions were sprayed using a pressure atomizer with LabVIEW \footnote{
{Laboratory Virtual Instrument Engineering Workbench (LabVIEW) is a system-design platform and development environment for a visual programming language from National Instruments. LabVIEW is commonly used for data acquisition, instrument control, and industrial automation on a variety of operating systems (OSs), including Microsoft Windows as well as various versions of Unix, Linux, and macOS.}
} 
software controlled system ~\cite{LABVIEW} in pulses of 100 ms at about 3 Hz frequency.

The substrates were kept at $\sim$150$~^{0}C$ and rotated by magnetic stirrer; the distance between the atomizer nozzle and the substrate was around 10 mm. The samples were coated with about 250 shots/cm$^2$. In a previous study of photoemission as function of the number of shots we observed that the photocurrent was increasing with increasing number of shots and saturated above 100 shots/cm$^2$. 

\subsection{QE measurements of different ND powders in different Gas Mixtures}

\begin{figure}
\begin{minipage}[c]{1\textwidth}
    \includegraphics[width=\textwidth]{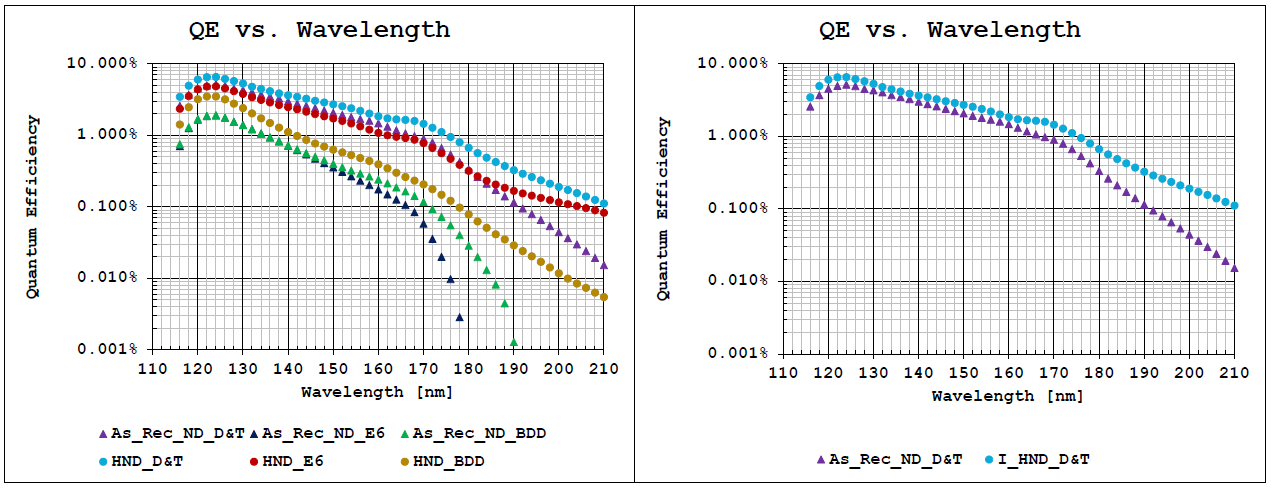}
\end{minipage}\hfill
\begin{minipage}[c]{1\textwidth}
        \caption{Left: QE as a function of wavelength for different types of ND and H-ND coated PCB substrates. Right: QE vs wavelength for the two types of photo cathodes providing highest values.}
        \label{fig:qeallcomp}
    \end{minipage}
\end{figure}

Test substrates have been coated with the six types of powders and their QE has been measured for each of them, for different wavelengths, with fixed electric field (0.2 kV/cm)  applied over the surface in vacuum as shown in Fig.\ref{fig:qeallcomp}-Left. The ND and H-ND powders from D\&T provided the highest QE for all wavelengths and are shown separately in Fig.~\ref{fig:qeallcomp}-Right.

\begin{figure}
\begin{minipage}[c]{1\textwidth}
    \includegraphics[width=\textwidth]{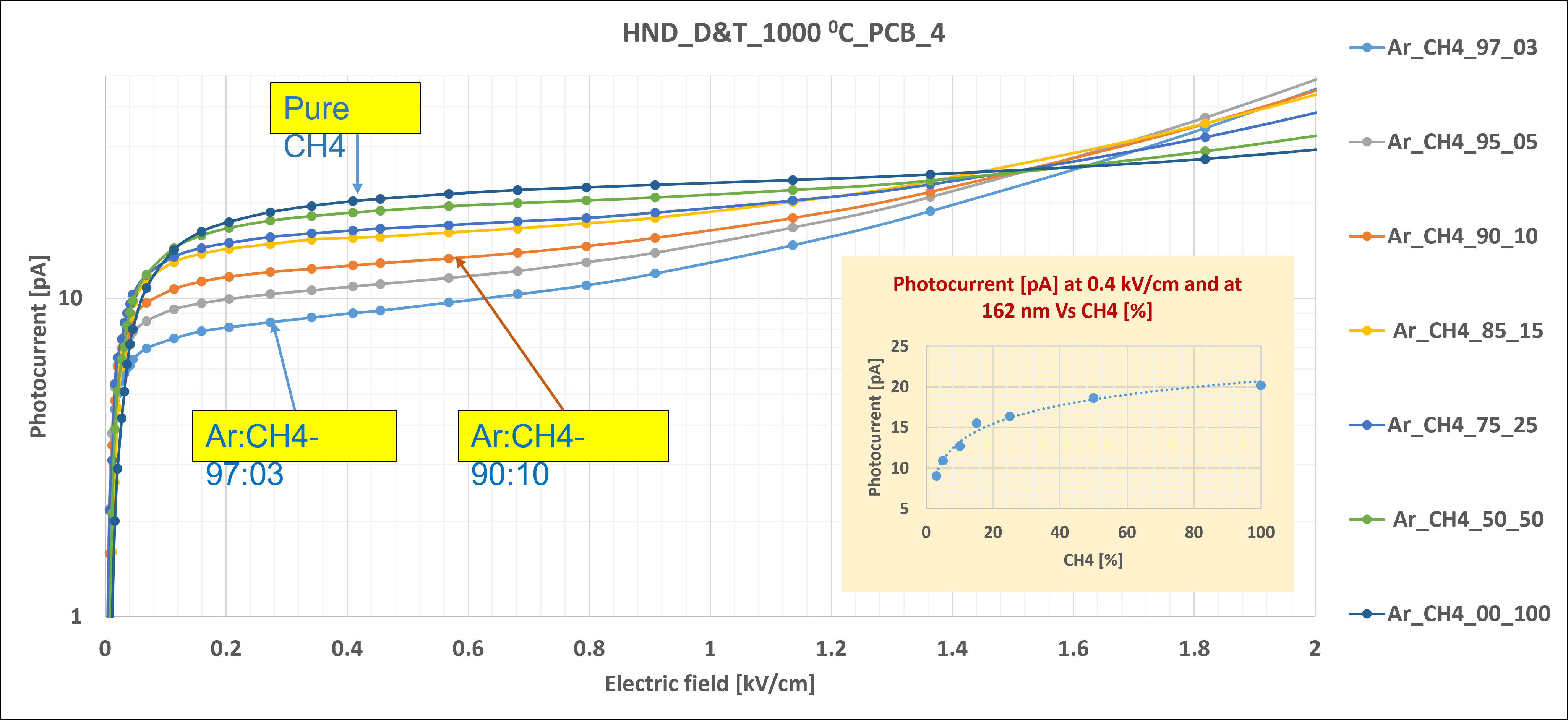}
\end{minipage}\hfill
\begin{minipage}[c]{1\textwidth}
        \caption{Photocurrent as a function of applied electric field over the photo-cathode surface for different Ar:$CH_{4}$ gas mixtures at 160 nm wavelength. The photocurrent values for 0.4 kV/cm as function of the fraction of CH$_4$ are presented in the inserted colored canvas.}
        \label{fig:PhEmFull}
    \end{minipage}
\end{figure}

The hydrogenated D\&T ND powder has the highest quantum efficiency for all the wavelength in vacuum. Systematic measurements have been performed at the peak wavelength 160 nm for different electric fields in different Ar:$CH_{4}$ gas mixtures.
From Fig.~\ref{fig:PhEmFull} we can see that the photoemission in Ar rich mixtures and at lower values of applied electric field the measured photo current is lower compared to that of $CH_{4}$ rich mixtures. That is a clear effect of the Ar high photon back scattering cross section.
\par
For mixtures with low percentage of $CH_{4}$ the effect of gas multiplication is visible for electric field values larger than 0.5 kV/cm and above $\sim 1.5 kV.cm^{-1}$ the Ar rich mixtures provide larger total measured photocurrent values compared to $CH_{4}$ rich gas mixtures because the gas multiplication effect becomes dominant.
\par
The photocurrent values for 0.4 kV/cm as function of the fraction of CH$_4$ are presented in the colored canvas inserted in Fig\ref{fig:PhEmFull}. With respect to the photocurrent provided by the pure $CH_{4}$ gas, the Ar/$CH_{4}$ 50/50 mixture shows a 9\% lower value, the Ar/$CH_{4}$ 75/25 a 24\% lower value and the Ar/$CH_{4}$ 97/03 a 55\% lower value.

\subsection{THGEM characterization}

\par
THGEMs are robust gaseous electron multipliers based on GEM, principle scaling the geometrical parameters. 
They are obtained via standard PCB drilling and etching processes. The 35~$\mu$m copper layer is coated with $\approx$5~$\mu$m of Ni, followed by 200~nm Au. The THGEMs used for our studies have an active area of $30\times30~mm^{2}$ with a  hole diameter of  0.4~mm, a pitch of 0.8~mm, a thickness of 470~$\mu$m and no rim around the holes.
\par

\begin{figure}
\begin{minipage}[c]{1\textwidth}
    \includegraphics[width=\textwidth]{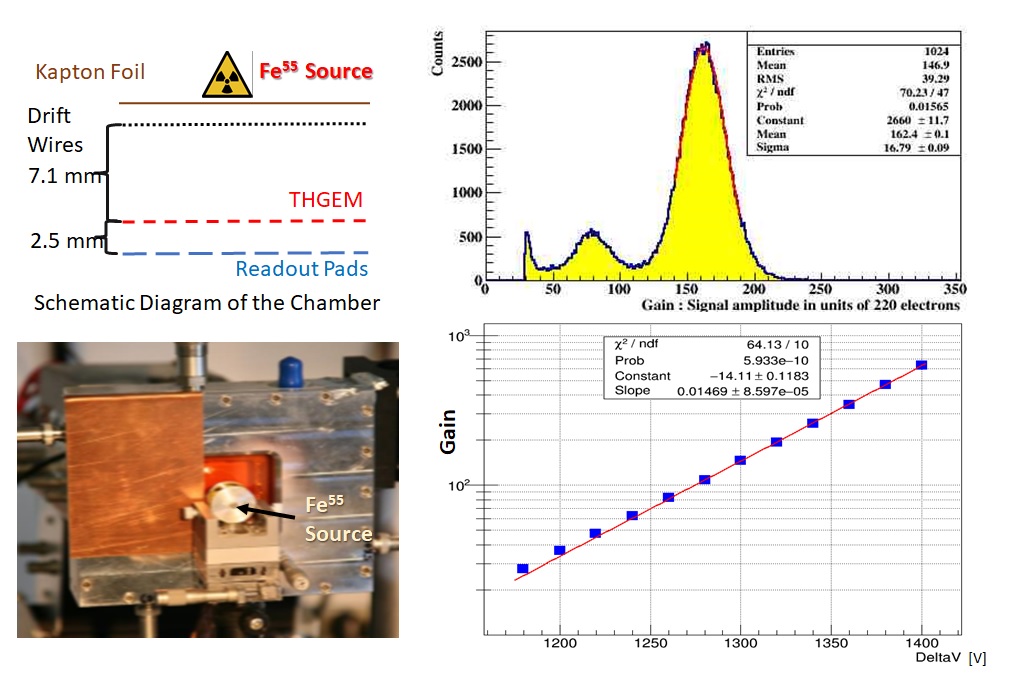}
\end{minipage}\hfill
\begin{minipage}[c]{0.98\textwidth}
        \caption{Top-left panel: Schematic of detector assembly. Bottom-left panel: Detector illuminated with an ${}^{55}Fe$ X-ray source. Top-right panel: Typical ${}^{55}Fe$ X-ray spectrum obtained in $Ar-CO_{2}~(70\%-30\%)$ gas mixture when the applied voltages at drift, top and bottom of THGEM are -2520 V, -1720 V and -500 V, respectively, while the anode is at ground. Bottom-right panel: Gain dependence of the THGEM versus the applied voltage.}
        \label{fig:Schematic_of_Detector_setup}
    \end{minipage}
\end{figure}

Each  THGEM is characterized in the setup sketched in figure  ~\ref{fig:Schematic_of_Detector_setup}. 
A plane of drift wires above it and a segmented readout anode plane, both properly biased, provide the drift and induction field respectively. The detector is operated with various gas mixtures, all including Ar. The electrons from $^{55}Fe$ converted by Ar are collected and multiplied in the hole region of the THGEM. The electron avalanche generated in the multiplication process, while drifting towards the anode,  induces 
the detected signal. 

All THGEMs have been characterized using a Ar/CO$_2$ 70/30 gas mixture and a  ${}^{55}Fe$  X-ray source at INFN Trieste before applying the coating procedures with the goal of performing comparative studies after coating them with VUV sensitive films. Environmental conditions (gas pressure and temperature) were registered and used for the corrected effective gain determination.
The THGEM gain response has a variation in time related to the effect of charging up of
the dielectric surfaces: a gain evolution study of about three days has been performed for each THGEM sample.
The same characterization has been repeated after coating.
\par 
 A typical ${}^{55}Fe$ X-ray spectrum obtained in  $Ar-CO_{2}~(70\%-30\%)$ gas  mixture is shown in   figure~\ref{fig:Schematic_of_Detector_setup}, top-right panel. The bottom right panel of figure \ref{fig:Schematic_of_Detector_setup} shows the gain dependence of THGEM versus the voltages applied between the two faces.

\begin{figure}
\begin{minipage}[c]{1\textwidth}
    \includegraphics[width=\textwidth]{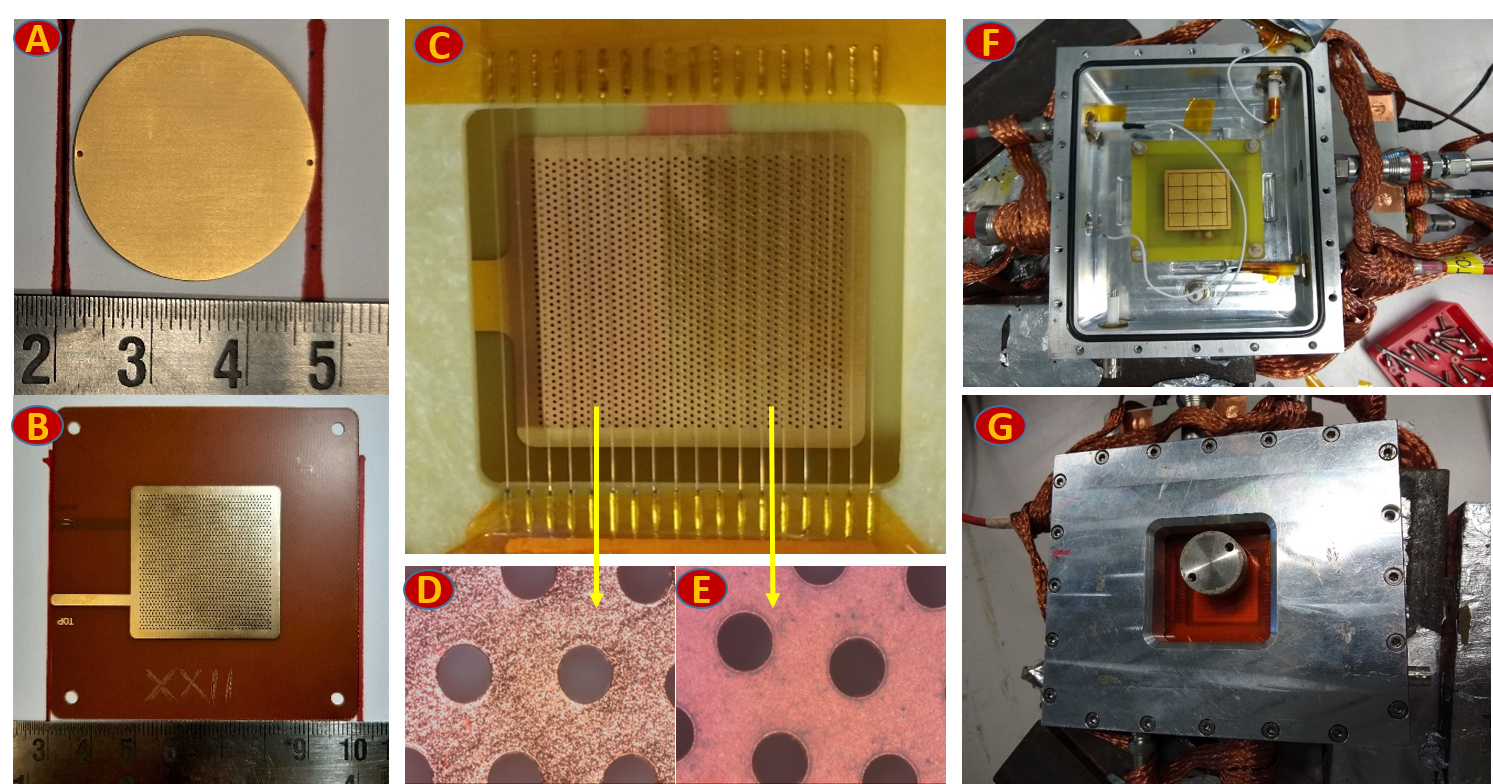}
\end{minipage}\hfill
\begin{minipage}[c]{0.98\textwidth}
        \caption{ (A) Au\_PCB of 1 inch diameter substrate used for the QE measurement. (B)  Uncoated THGEM of active area 30~mm$\times$30~mm. (C) Half uncoated and half coated THGEM, mounted into the test chamber and zoomed view of the both coated (D) and uncoated (E) part. (F) test chamber with readout pad where the THGEMs are tested. (G) The test chamber after installation of a THGEM, illuminated by an  ${}^{55}Fe$ X-ray source.}
        \label{fig:Detector_Images}
    \end{minipage}
\end{figure}

THGEMs have been coated in Bari either with as-received ND powder, or with H-ND powder.  
To allow direct comparison, for some THGEMs only half of the active surface has been coated,
leaving the other half uncoated. The same coating procedure (and the same amount of ND per unit surface) has been applied to test substrates and THGEMs.  
 \par
 Images of the coated substrates and the setup for the characterization are provided in Fig.~{\ref{fig:Detector_Images}}.


\begin{figure}
\includegraphics[width=0.56\linewidth]{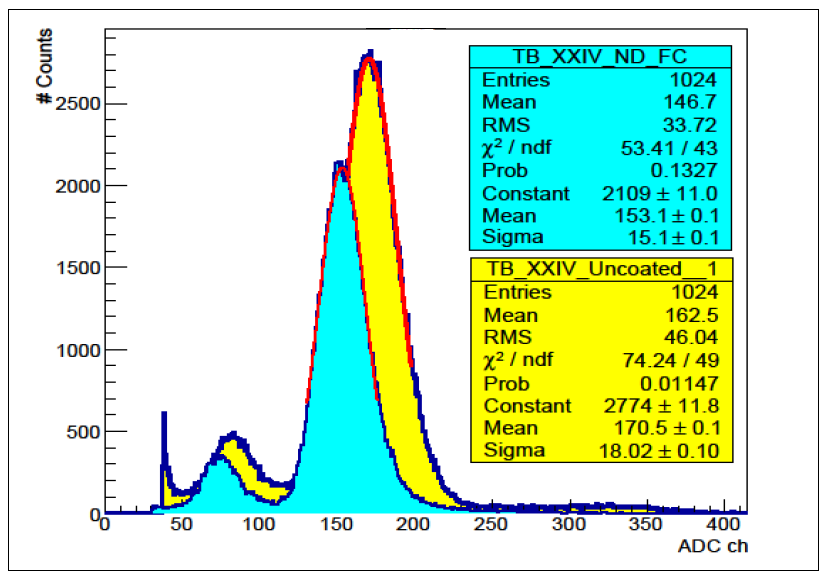}
\includegraphics[width=.43\linewidth]{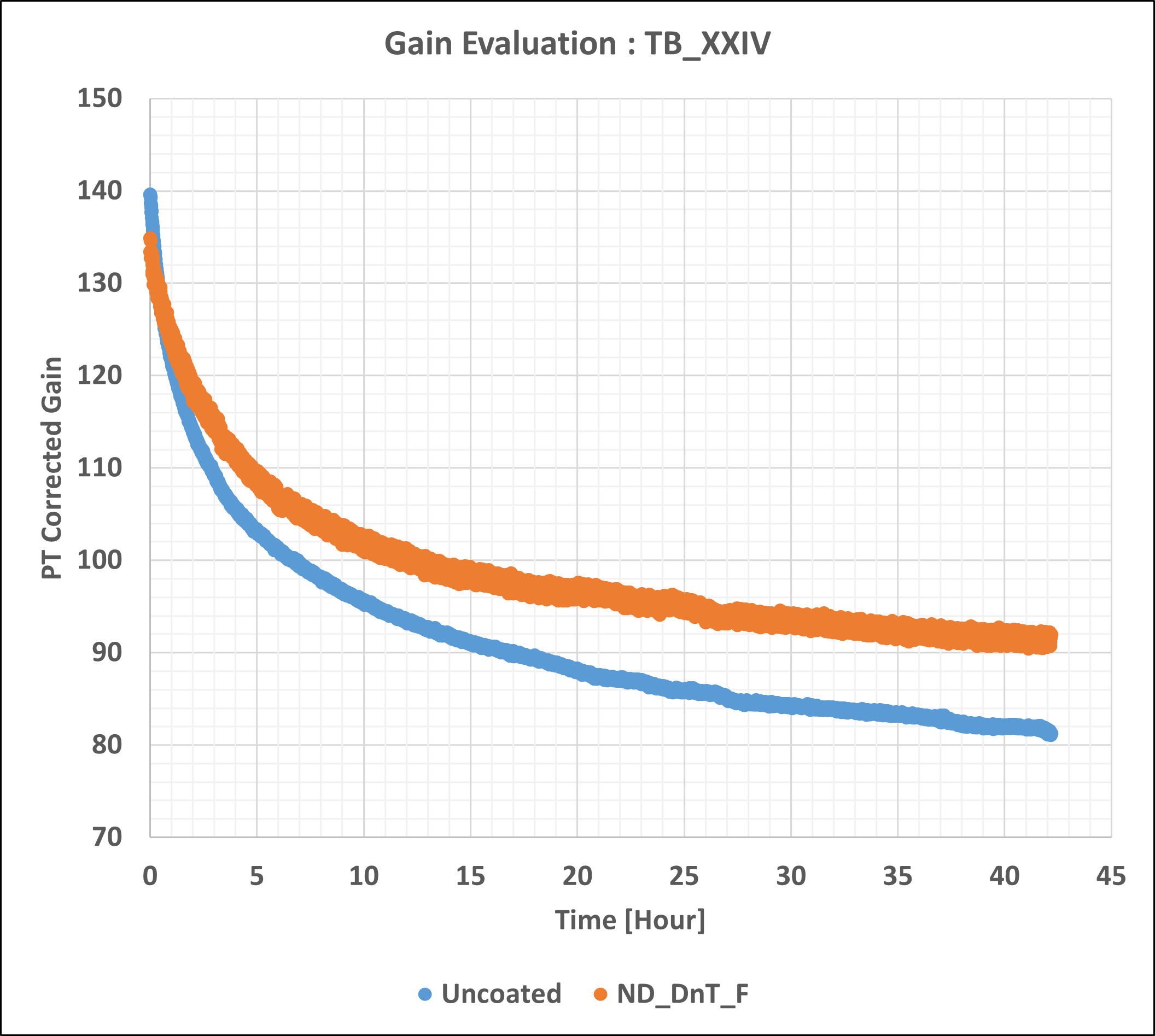}
\caption{Left: typical ${}^{55}Fe$ X-ray spectra obtained from a same THGEM-XXIV  before coating and after fully coated ND powder in  $Ar-CO_{2} (70\%-30\%)$ gas  mixture.  The voltages applied  to drift, top and bottom of the THGEM electrodes are 2250 V, 1750 V and 500 V respectively, while the anode is kept at ground.
Right: gain evolution due to charging-up of the THGEM measured before (blue) and after (orange) coating; PT correction has been applied.
}
\label{fig:ND_THGEM_Spectra}
\end{figure}

\par 
Some of fully characterized THGEMs with different coatings are listed below: 

\begin{table}[!ht]
    \centering
    \begin{tabular}{|l|l|l|l|}
    \hline
        Uncoated THGEMs & Max $\Delta$V THGEM & Coated THGEMs & Max $\Delta$V THGEM  \\ \hline
        TB\_XV & 1325 & H-ND\_BDD\_H & 1325 \\ \hline
        TB\_XVI & 1375 & ND\_E6\_H & 1400 \\ \hline
        TB\_XVII & 1350 & ND\_BDD\_H & 1325 \\ \hline
        TB\_XXIV & 1375 & ND\_D\&T\_F & 1325 \\ \hline
    \end{tabular}
    \caption{Maximum stable voltage of few uncoated and coated THGEMs in  $Ar/CO_{2} 70/30$ gas  mixture. }
    \label{tab:TableDelVThGEM}
\end{table}

The first column of Tab.\ref{tab:TableDelVThGEM} gives the THGEM labels, the second the maximum voltage bias at which the THGEM was operating stably, namely without discharging; in the third the coating material for each THGEM is listed and in the fourth the maximum stable voltage after coating. The electrical stability of THGEMs does not seem to be systematically affected by the coating layers. The variations are compatible with the changes in environmental conditions.
\par
Typical examples of amplitude distributions for ${}^{55}Fe$ signals from an uncoated (yellow) and ND coated (light blue) THGEM are shown superimposed in the left part of Fig. ~\ref{fig:ND_THGEM_Spectra}. The amplitude spectra are presented before application of the correction for the different pressure and temperature conditions. The effective gain is extracted from the mean of the Gaussian fit of the main peak of the spectrum.
To study the charging-up response, a continuous set of short ($\sim$ 30 sec.) acquisition runs
over few days is performed: in the right part of Fig.~\ref{fig:ND_THGEM_Spectra} the charging-up curve is presented for the same THGEM before (blue) and after (orange) coating; for this comparison the effective gain values have been corrected for pressure and temperature.
The difference seen in the charging up variation rate is partly due to the change in the ${}^{55}Fe$ source activity. 
\par
\begin{figure}
\includegraphics[width=.98\linewidth]{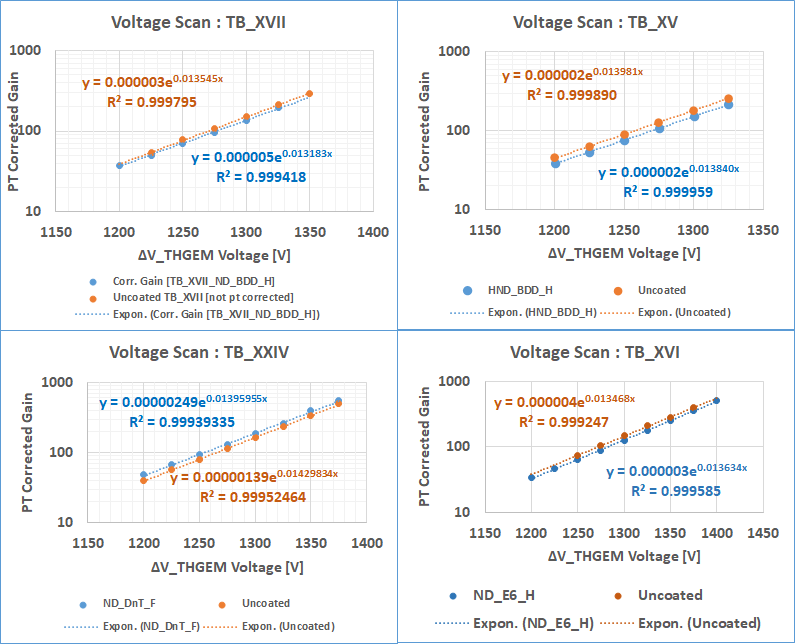}
\caption {Comparison between corrected effective gain versus applied voltage across the THGEM electrodes before and after coating for four THGEMs with different type of coating.}
\label{fig:HV_scan}
\end{figure}

The values of the effective gain corrected for pressure and temperature as a function of the bias voltage applied to the THGEMs are presented in Fig.~\ref{fig:HV_scan} for the four THGEM samples of Tab.~\ref{tab:TableDelVThGEM}, before and after coating. 
The characterization of THGEMs without and with various ND coating provides interesting indications. The gain values are found to be similar in all cases: the THGEM response as electron multiplier is not seriously modified by any of the ND and H-ND coatings.

\section{Conclusion}
The ongoing R\&D programme to explore the H-ND photocathode response in the VUV range for future gaseous PD application has provided encouraging results.
Spectral photoemission current measurements in the VUV range between 116 nm and 210 nm in vacuum and various gas mixtures were performed for ND and H-ND samples. 
Among recently tested ND powders, the nanodiamond provided by D\&T S.r.l. showed the highest QE values. 
The first systematic photoemission measurements from ND and H-ND photocathodes in various $Ar-CH_{4}$ gas mixtures is presented.
Photoemission currents measured at 160 nm as a function of the electric field, in various $Ar-CH_{4}$ compositions, allowed to distinguish between two different voltage regimes. At lower field values the important role of $CH_{4}$ in reducing the photoelectron backscattering is highlighted, providing a factor of two in photoconversion probability between $Ar$-rich mixtures and pure $CH_{4}$.
 At higher voltages the gas multiplication dominates, in particular for the gas mixtures with high $Ar$ percentage. 
\par 
THGEM samples  coated with different types of photosensitive layers of ND and H-ND, from D\&T, E6 and BDD  have been studied with extensive characterization. 
It is observed that these ND and H-ND coated THGEMS sustain more than 1325 V bias across top and bottom surfaces in all cases.
The response of THGEMs as electron multipliers is not significantly modified by the ND or H-ND coating.
The results of this first exploratory study, suggesting a full compatibility of H-ND with THGEMs encourage to proceed in the development of gaseous PDs with a photocathode overcoming the limitations of CsI.

\section{Acknowledgment}
This R\&D activity is partially supported by
\begin{itemize}
\item 
EU Horizon 2020 research and innovation programme, STRONG-2020 project, under grant agreement No 824093;
\item 
the Program Detector Generic R\&D for an Electron Ion Collider by Brookhaven National Laboratory, in association with Jefferson Lab and the DOE Office of Nuclear Physics.
\end{itemize}


\begin{thebibliography}{99}




\bibitem{EPIC}
R. Abdul Khalek et al., \emph{Science Requirements and Detector Concepts for the Electron-Ion Collider: EIC Yellow Report}, 
Nuclear Physics A \textbf{1026}, (2022), 122447

\bibitem{EIC}
A. Accardi et al., \emph{Electron-Ion Collider: The next QCD frontier}, Eur. Phys. J. A \textbf{52} (2016) 268.

\bibitem{Frank:1937fk}
I.~Frank and I.~Tamm,
\emph{``Coherent visible radiation of fast electrons passing through matter,''}
Compt. Rend. Acad. Sci. URSS \textbf{14}, \textbf{no.3},  (1937) 109-114.

\bibitem{MPGD}
Fabio Sauli et al., \emph{GASEOUS DETECTORS HANDBOOK
}, http://fabio.home.cern.ch/fabio/handbook.html

\bibitem{windowless-RICH}
M. Blatnik et al., \emph{Performance of a Quintuple-GEM Based RICH Detector Prototype}, IEEE NS \textbf{62} (2015) 3256.

\bibitem{PM18}
J. Agarwala et al., \emph{The MPGD-based photon detectors for the upgrade of COMPASS RICH-1 and beyond}, Nucl. Instr. and Meth. A \textbf{936}  (2019) 416.

\bibitem{thgem}
L. Periale et al., \emph{Detection of the primary scintillation light from dense Ar, and Xe with novel photosensitive gaseous detectors}, Nucl. Instr. and Meth. A \textbf{478}  (2002) 377; \newline
P. Jeanneret, \emph{Time Projection Chambers and detection of neutrinos, PhD thesis}, Neuchatel University, 2001; \newline
P.S. Barbeau et al, \emph{Toward coherent neutrino detection using low-background micropattern gas detectors} IEEE \textbf{NS-50} (2003) 1285; \newline
R. Chechik et al, \emph{Thick GEM-like hole multipliers: properties and possible applications}, Nucl. Instr. and Meth. A \textbf{535} (2004) 303.

\bibitem{mm}
Y. Giomataris et al., \emph{MICROMEGAS: a high-granularity positionsensitive gaseous detector for high particle-flux environments}, Nucl. Instr. and Meth. A \textbf{376} (1996) 29.

\bibitem{RD26}
The RD26 Collaboration, \emph{RD26 status reports: CERN/DRDC} \textbf{93-36} (1993), CERN/DRDC \textbf{94-49} (194), CERN/DRDC (1996).

\bibitem{NIMA_695_2012_279}
Triloki n, B.Dutta,B.K.Singh, \emph{Influence of humidity on the photoemission properties and surface morphology of cesium iodide photocathode},  Nucl. Instr. and Meth. A, \textbf{695} (2012) 279.


\bibitem{NIMA_574_2007_28}
H.Hoedlmoser et al., \emph{Long term performance and ageing of CsI photocathodes for the ALICE/HMPID detector}, Nucl. Instr. and Meth. A, \textbf{574} (2007) 28.

\bibitem{ibf-blocking}
A. Bondar et al., \emph{Study of ion feedback in multi-GEM structures},  Nucl. Instr. and Meth. A \textbf{496} (2003) 325; \newline
A. Breskin et al., \emph{Sealed GEM photomultiplier with a CsI photocathode: ion feedback and ageing}, Nucl. Instr. and Meth. A \textbf{478} (2002) 225; \newline
J.F.C.A. Veloso et al., \emph{A proposed new microstructure for gas radiation detectors: The microhole and strip plate}, Rev. Sc. Instr. \textbf{71} (2000) 2371; \newline
A.V. Lyashenko et al., \emph{Further progress in ion back-flow reduction with patterned gaseous hole - multipliers},JINST \textbf{2} (2007) P08004;\newline
A.V. Lyashenko et al., \emph{Efficient ion blocking in gaseous detectors and its application to gas-avalanche photomultipliers sensitive in the visible-light range}, Nucl. Instr. and Meth. A \textbf{598} (2009) 116;\newline
M. Alexeev et al., \emph{Ion backflow in thick GEM-based detectors of single photons}, JINST \textbf{8} (2013) P01021. \newline
M. Bari et al., \emph{RHIP, a Radio-controlled High-Voltage Insulated Picoammeter and its usage in studying ion backflow in MPGD-based photon detectors}; PoS \textbf{MPGD2017} 068; \newline
M.M.~Aggarwal et al, (ALICE TPC Collaboration), \emph{Particle identification studies with a full-size 4-GEM prototype for the ALICE TPC upgrade}, Nucl. Instr. and Meth. A \textbf{903} (2018) 215.

\bibitem{JAP_77_1995_2138} A.Buzulutskov, A.Breskin, R.Chechik, \emph{Field enhancement of the photoelectric and secondary electron emission from CsI} ,Journal of Applied Physics \textbf{77} (1995) 2138. 

\bibitem{NDRep-1}
L. Velardi, A. Valentini, G. Cicala, \emph{Highly efficient and stable UV photocathode based on nanodiamond particles}, Appl. Phys. Lett. \textbf{108} (2016) 083503-1-5.


\bibitem{NDreport}
A. Valentini, D. Melisi, G. De Pascali, G. Cicala, L. Velardi, A. Massaro, \emph{High-efficiency nanodiamond-based ultraviolet photocathodes}, 30-03-2017 Patent n. \textbf{WO 2017/051318 A9}; International Patent n. \textbf{PCT/IB2016/055616} of September 21, 2016; National Patent Italia - n. 102015000053374 del 21 Settembre 2015, Istituto Nazionale di Fisica Nucleare e Consiglio Nazionale delle Ricerche.


\bibitem{NIMA_438_1999_94}
H. Rabus, et. al., \emph{Quantum efficiency of cesium iodide photocathodes in the 120-220 nm spectral range traceable to a primary detector standard}, Instr. and Meth. A , \textbf{438} (1999) 94.

\bibitem{JPCM_21_2009_16}
F Silva, K Hassouni, X Bonnin and A Gicquel “Microwave engineering of plasma-assisted CVD reactors for diamond deposition” J. Phys.: Condens. Matter 21 (2009) 364202 (16pp).

\bibitem{Pat_Spec_5_501_740}
Besen M M, Sevillano E and Smith D K 1996 Patent Specification 5,501,740.

\bibitem{coating-I}
G. Cicala, A. Massaro, L. Velardi, G. S. Senesi, A. Valentini, \emph{Self-Assembled Pillar-Like Structures in Nanodiamond Layers by Pulsed Spray Technique}, ACS Appl. Mater. Interfaces \textbf{6} (2014) 21101.

\bibitem{coating-II}
L. Velardi, A. Valentini, G. Cicala, \emph{UV photocathodes based on nanodiamond particles: effect of carbon hybridization on the efficiency}, Diam. Relat. Mater. \textbf{76} (2017) 1-8.

\bibitem{LABVIEW}
\emph{https://www.ni.com/en-in/support/downloads/software-products/download.labview.html\#460283}

\bibitem{asset}
$https://indico.cern.ch/event/872501/contributions/3726017/attachments/1985809/3308869/Marta\_Lisowska\_-\_RD51\_Mini\_Week\_-\_Asset\_photocathode\_characterisation\_device.pdf$
\end{thebibliography}
\end{document}